\documentclass[aps,amsmath,twocolumn]{revtex4}

\usepackage{amsmath}
\usepackage{graphicx}

\begin{document}

\title{Superluminal light propagation in a
bi-chromatically Raman-driven and Doppler-broadened N-type 4-level
atomic system}
\author{Bakht Amin Bacha}
\affiliation{Department of Physics, Hazara University, Pakistan}
\author{Fazal Ghafoor}
\affiliation{Department of Physics, COMSATS Institute of
Information Technology, Islamabad, Pakistan}
\author{Iftikhar Ahmad}
\affiliation{Department of Physics, Malakand University Chakdara
Dir (L), Pakistan}

\begin{abstract}
We investigate the behavior of fast light pulse propagation in an
N-type Doppler-broadened 4-level atomic system using double Raman
gain processes. This system displays novel and interesting results
of two controllable pairs of the double gain lines profile with a
control field. The detailed physics of the processes are explored
having multiple controllable anomalous regions in the medium. In
this set up, the system exhibits significant enhancement in the
probing Gaussian pulse through the medium as compared with Ref.
[L. J. Wang, A. Kuzmich, and A. Dogariu, Nature \textbf{406},
227(2000)]. The advance time of the retrieved Gaussian pulse is
always greater than the advance time studied in the above said
experiment. We analyzed that the pulse propagating through the
medium with larger negative group index, $7.32\times10^8$, leaves
the medium almost undistorted and sooner by time $76.12 \ ms$ than
the pulse which leaves the medium of Wang \emph{et al.}. The
Gaussian pulse always remains almost undistorted at output due to
lossless characteristic of the medium. We also underlined the ways
to suppress decoherences generated by the Doppler-broadening in
the system. The limitations of the recently developed applications
require to explore mechanisms for ultimate speed of a superluminal
light pulse. In this connection, the proposed scheme may be
helpful and can be easily adjusted with the current technology.
\end{abstract}

\maketitle

\section{Introduction}
The purpose of research on the manipulation of light group
velocity from
subluminal to superluminal is not only to study a novel state of matter \cite%
{Kang} but also to explore its applications in the areas of
optical memory, optical computing and optical communication
\cite{Tseng,Dahan}. In most of these studies coherent fields are
used to control the optical properties of a medium. Consequently,
this has led to many remarkable results such as enhanced nonlinear
optical effects \cite{Tewari,Harris}, Electromagnetically Induced
Transparency (EIT) \cite{Harris01}, Lasing Without Inversion (LWI)
\cite {Kocha,Kocha01,Agarwal}, ultra-slow light
\cite{Hau,Kash,Budker,Turukhin,Ku}, storage and retrieval of
optical pulses \cite{Philiops}, and many others
\cite{Zibrov,Kang01,Deng,Stenner}.

The pioneer experimental work of superluminal propagation of light
pulse are credited to Chu and Wong \cite{Chu} in the optical
domain using a correlation technique. The credit also goes to
Segard and Macke \cite{Segal} for the demonstration of
superluminality in microwave domain with a direct detection of the
transmitted intensity of suitably shape millimeter wave pulses
through a linear molecular absorber. In both the cases the
propagated pulse is less distorted. These results successfully
verifies the prediction of Garrett and McCumber \cite{Garrett}. A
remarkable control of superluminal group velocity of light in a
system has also been demonstrated experimentally in Ref.
\cite{Chiao}. However, using saturation effect the pulse advances
in time through an amplifying medium is studied in Ref.
\cite{Basov}. More recently, gain-assisted anomalous dispersion
which leads to superluminal light propagation has also been
demonstrated in Refs. \cite{Steinberg,Wang}. The beating of pump
and probe field which generates coherent population pulsation in a
medium exhibits superluminality \cite{Bigelow}. A large number of
studies exist in literature in different contexts which describe
superluminal light pulse propagation, for example, see Refs. \cite
{FCARRELIO2005,Bigelow,AM1999,KKIM2003,HKANG2003,FXIAO2004,tj04}.
However, in this paper we restrict ourselves to the investigation
of a viably gain-assisted atomic system for a superluminality
Gaussian light pulse and its enhancement during propagation in the
dispersion medium.

Wang \textit{et al.} \cite{Wang} demonstrated superluminal light
pulse propagation for the first time using region of lossless
anomalous dispersion between the two closely spaced gain lines in
a double Raman gain medium. They observed that the dispersion
properties of the medium can be manipulated to have negative group
index for the superluminality. Moti \textit{et al.} utilized this
approach for temporal cloaking \cite{Fridman2012} using the
concept of spatial cloaking \cite{Leon2006,peny2006} in the
temporal domain. It is the manipulation of positive and negative
group indices due to which the probe light pulse propagates around
an object in such a way that creates a hole both in time and space
windows. This behavior enabled us to hide a physical event from
any physical observer. The future aims of these and their related
experiments are to explore spatio-temporal cloaking for making
history of a physical event completely hidden. However, due to the
limitations of current technology, the present problem in these
experiments is the less gap creation. Therefore, it requires to
increase the gap to microsecond and to millisecond \cite{RB2012}.
Obviously, we ultimately high group velocity for larger gap
creation without significant distortion of the pulse shape at the
output of the medium. To the best of our knowledge, less interest
has been shown by researchers in this area. Furthermore, following
the experiment of temporal cloaking, Glasser \emph{et al}.
\cite{RYANT2012} utilized the negative group velocity of light for
the measurement of images. However, in their experiment, as they
discussed therein, we need an ultimate speed of the light pulse to
get improve the quality of the images. According to these facts
and may be many more, it is necessary to explore mechanism for an
ultimate speed of light pulse with a minimum distortion of a light
pulse at the output \cite{Leonh2009,Fridman2012,Mc2011}.

In this paper we propose two-paired gain assisted model for
demonstration of enhancement of superluminality of a Gaussian
light pulse. In this model, we initially couple two pump fields
appropriately in the atomic system to have initially double Raman
gain lines feature. We also couple another control field with a
fourth energy level to provide a source to change the single pair
of the gain lines to two controllable pairs. This behavior
provides multiple controllable anomalous regions in the dispersion
medium. Furthermore, we also consider the medium Doppler
broadening effect in the system due to high temperature for a
maximum Doppler width. This broadening sensitivity is in complete
contrast with the experimental scheme of Wang \textit{et al.}
\cite{Wang}, where the induction of Doppler effect in their system
creates problem of singularities and consequently converts the
gain processes to absorption. In fact, their experiment, is
Doppler-broadening insensitive and limits the superluminal effect
of the propagating pulse. The control coupling field along with
the two pump fields and the probe field makes our proposed system
Doppler-broadened. The appropriate counter propagating fields
provide maximum Doppler effect in the system to maximize the speed
of the superluminal pulse to its ultimate value. The
superluminality of a Doppler-broadened system can be remarkably
enhanced to its highest values if inhomogeneously broadened
solid-state system is considered where the densities of the atomic
medium are very large.

In principle, the lossless character of the Wang \emph{et al.}
scheme provides a ground to demonstrate the superluminal effect in
a laboratory along with almost undistorted pulse shape of light at
the output of the medium. Nevertheless, our proposed scheme not
only exhibits the same lossless behavior but with the advantages
of multiple anomalous regions. In addition, the Doppler broadening
sensitivity of the system provided by the control field enhances
the superluminality tremendously. Quantitatively, the analytical
results for the output pulse shape is calculated to fourth order
perturbation limit in the group index. It is shown that even with
a significant enhancement in the group velocity of a Gaussian
pulse, its shape remains undistorted. This behavior of significant
enhanced superluminal light pulse with its almost undistorted
shape received at the output, is remarkable. Therefore, it may be
of interest for researchers to demonstrate these mechanisms in a
laboratory.
\section{Model and equation}
\begin{figure}[t]
\centering
\includegraphics[width=3.5in]{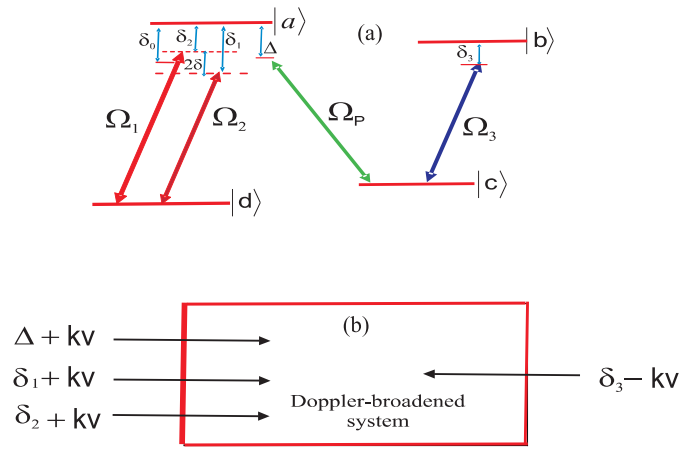}
\caption{(a) Schematics of  the atomic system. (b)
Doppler-broadened system} \label{figure1}
\end{figure}
We consider a 4-level double-Raman atomic-system in N-type
configuration driven by two coherent pump fields, a probe and a
control field, appropriately [see Fig. 1]. The two lower levels
$\left\vert d\right\rangle $ and $\left\vert c\right\rangle $ are
coupled with the upper level $ \left\vert a\right\rangle $ by the
two pump fields of Rabi frequencies $\Omega _{1}$, $\Omega _{2}$
and with a probe field of Rabi frequency $\Omega _{p}$,
respectively. Meanwhile, the level $\left\vert c\right\rangle$, is
also coupled with another fourth level $\left\vert
b\right\rangle$, by a control field of Rabi frequency $ \Omega
_{3}$. Further, to present the model and equations of motion for
this system we proceed with the following interaction picture
Hamiltonian in the dipole and rotating wave approximations:
\begin{eqnarray}
H(t) &=&-\frac{\hbar }{2}(\Omega _{1}e^{-i\delta _{1}t}+\Omega
_{2}e^{-i\delta _{2}t})\left\vert a\right\rangle \left\langle
d\right\vert
\notag \\
&&-\frac{\hbar }{2}\Omega _{3}e^{-i\delta _{3}t}\left\vert
b\right\rangle \left\langle c\right\vert -\frac{\hbar }{2}\Omega
_{p}e^{-i\Delta t}\left\vert a\right\rangle \left\langle
c\right\vert +H.c.
\end{eqnarray}%
where, $\omega_1=\omega_{ad}\pm\delta_1$,
$\omega_2=\omega_{ad}\pm\delta_2$, $
\omega_3=\omega_{bc}\pm\delta_3$ and
$\omega_p=\omega_{ac}\pm\Delta$. In the Eq. (1) the chosen
detuning parameters are $\delta _{1}=\delta _{0}-\delta ,$ and
$\delta _{2}=\delta _{0}+\delta $. Further, the detuning $\delta
=\frac{\delta _{2}-\delta_{1}}{2}$, appears for effective
frequency of the two pump fields while $\delta _{0}=\delta
_{1}+\delta _{2}/2$, is the average detuning of the two pump
fields from the atomic transition. Next, to find out the equations
of motion we use the following general form of density matrix
equation:
\begin{equation}
\frac{d\rho }{dt}=-\frac{i}{\hbar }[H(t),\rho ]+\Lambda \rho ,
\end{equation}%
where, $\Lambda \rho $, is the damping part of our system. Using
the above density matrix equation and following the transformation
relations for slowly and fast varying amplitudes we calculated
rate equations for the proposed system. These rate equations are
listed in the Appendix-A. We assume the phase factor associated
with the two modes of the coherent pump fields constant following
the series of papers of Wang and his co-workers
\cite{Wang,Cao,Wangpra,Wangprl}. In Ref.
\cite{Wang,Wangpra,Wangprl} they presented their theoretical
results of dispersion for the constant phase. Further, in Ref.
\cite{Cao} they presented detailed conceptual interpretation of
the phases associated with the driving fields frequencies using
wave nature characteristics of the classical Maxwell's equations
for a light pulse propagating through the transparent region of
the medium. Due to the negative group index and its associated
negative time delay the phases of the driving fields get re-shape
to have zero phase throughout the transparent medium. It gets back
the same value of the phases of the associated entering pulses
after the exit. These facts are discussed in detail using
classical Maxwell's equations for wave nature of a light pulse
propagation before medium, in the medium and after the medium,
respectively. Furthermore, the detailed analysis carried in Ref.
\cite{Cao} for the corresponding phase also agree with the
experimental results reported in \cite{Wang,Wangpra}. They also
given the detail procedure that how this phase can be calculated
and can be related with group index of the medium while using the
experiment for the superluminality [see Ref. \cite{Wangpra}]. In
fact, It is hard to calculate the dispersion of the propagating
pulses if the relative phases are considered time dependent.
Consequently, a numerical method is needed. Nevertheless, it is
not the case of the Wang \emph{et al.} results where they
calculated the two terms independently. This means constant
relative phase of value zero [see Ref. \cite{Wangpra}]. This is in
agreement to what they explained in Ref. \cite{Cao}. In our system
we handled the model in an exact way of defining the rate
equations while following the effective Hamiltonian of the system
for the two pump fields coupled with the same transition
simultaneously. Due to this reason a phase factor enters the
system as an interference factor in the dynamical equations. Under
these reasonable facts we evaluated the steady states solution for
$\widetilde{\rho}_{ac}$ as:
\begin{equation}
\overset{\sim }{\rho
}_{ac}=\frac{-i\Omega_p}{8}\sum^4_{i=1}P_i(\Delta),
 \end{equation}
where, $P_{j=1-4}(\Delta)$ are given in Appendix-B. This is a more
general expression and its corresponding susceptibility is reduced
to the susceptibility of wang \emph{et al.} in the next section
using some reasonable approximations in our system.
\section{susceptibility and group index}
The susceptibility is defined as a response function of medium due
to an applied electric field. The susceptibility of our driven
atomic system is calculated to the second order in the pump fields
and to the first order in the probe and control field. This is the
condition under which double Raman gain processes are studied. The
susceptibility of the system is calculated as:
\begin{eqnarray}
\chi=\frac{2N|\sigma_{ac}|^2\widetilde{\rho}_{ac}}{\epsilon_0\hbar\Omega_p}=\frac{-i3N \lambda^3\gamma}{32\pi^3}\sum^4_{i=1}P_i(\Delta),
\end{eqnarray}
where $N$ is the atomic number density of the medium. Furthermore,
$\lambda=2\pi c/\omega_{ac}$, and the Einstein coefficient
$A_{coff}=4\sigma_{ac}|^2\omega_{ac}^3/\epsilon_0\hbar c^3$ is
equal to $4\gamma$.

In principle, our system is more general than the famous
experimental system of Wang et al. while having significant
advantages. Consequently, under some reasonable approximations the
analytical results of the susceptibility of our system approaches
to the corresponding theoretical results calculated by Wang
\emph{et al.} Therefore, in our system if we assume
$\Gamma_{ad}=\Gamma_{bd}=0$, $\Omega_3=0$, $\Delta_3=0$,
$\Gamma_{ac}=\Gamma_{dc}=\Gamma$, $\delta_0\approx\Delta$,
$\delta/\delta_0,\Gamma/\delta_0<<1$,
$\delta_{1,2}=2\pi(\nu_{1,2}-\nu_{ad})$,
$\Delta=2\pi(\nu_p-\nu_{ac})$, $\nu_{ac}\approx\nu_{ad}$, we get
the required results i.e.,
\begin{eqnarray}
\chi_w=\frac{M_1}{\nu_p-\nu_1+i\gamma}+\frac{M_2}{\nu_p-\nu_2+i\gamma},
\end{eqnarray}
where $M_{j=1,2} =
N|\sigma_{ac}|^2|\Omega_{1,2}|^2/8\pi\epsilon_0\hbar\delta^2_{0}$
and $\sigma_{ac}$, is the dipole-moment between the levels
$\left|a\right\rangle$ and $\left|c\right\rangle$ and
$\gamma=\Gamma/2\pi$. Next, we consider the Doppler frequency
shifts induced by the atoms moving with a velocity $v$, relative
to the coherently driven pump fields, the control and the probe
field. The configuration of the counter propagating waves of these
driven fields through the atomic media is shown in the Fig. 1(b).
To explore the Doppler broadening effect in the system we replace
the detuning parameters such that $\delta_1=\delta_1+ k_1v$,
$\delta_2=\delta_2+k_2v$, $\Delta=\Delta+ k_pv$,
$\delta_3=\delta_3-k_3v$, in the Eq. (15) for the Eq. (18) of
susceptibility $\chi$, where $k_1=k_2=k_3=k_p=k$, and $v$, is the
atomic velocity of the medium. The susceptibility obtained is
given by:
\begin{eqnarray}
\chi(kv)=\frac{-3iN \lambda^3\gamma}{32\pi^3}f(kv,\Delta),
\end{eqnarray}
where $f(kv,\Delta)=\sum^4_{j=1}A_j(kv,\Delta)$. The parameters
$A_{j=1-4}(kv,\Delta)$ of the above equation are listed in the
Appendix-B. Now integrating $\chi(kv)$ over the velocity
distribution we have $\chi^{(d)}=\frac{1}{\sqrt{2\pi
V_D^2}}\int^\infty_{-\infty} \chi(kv) e^{-\frac{(kv)^2}{2V^2_D}}
d(kv).$ Here, $V_D$ is the Doppler width and is given by
$V_D=\sqrt{K_BT\omega^2Mc^2}$. Now, rewriting the above equation
in the following form
\begin{equation}
\chi(d)=\frac{-3iN
\lambda^3\gamma}{32\sqrt{2}\pi^{7/2}V_D}\int^\infty_{-\infty}f(kv,\Delta)e^{-(kv)^2/2V^2_D}d(kv).
\end{equation}
To keep the condition of atomic motion relative to the frequencies
of the driving fields more general we assumed the system
parameters modifiable due to flexible environment. Therefore we
consider the atomic velocity linear in the response function of
the medium of the system. respectively. Finally we write the group
index for our system when there is no Doppler effect as
\begin{equation}
N_g=1+2\pi
Re[\chi]+2\pi\omega_{ac}Re[\frac{\partial\chi}{\partial\Delta}],
\end{equation}
while
\begin{equation}
N_g(d)=1+2\pi
Re[\chi(d)]+2\pi\omega_{ac}Re[\frac{\partial\chi(d)}{\partial\Delta}],
\end{equation}
\begin{equation}
 \tau_d=\frac{L(N_g-1)}{c}
 \end{equation}
 These are our final results which will be analyzed and discussed in
details in the results and discussion section. $\tau_d$ is group
delay time when its value is positive. If the value of $\tau_d$ is
negative then is called advance group delay time($\tau_{ad}$)
\section{Distortion measurements}
A complex monochromatic wave-field of angular frequency $\omega$,
position $z$ and time $t$, is given by
$E(z,t)=\frac{1}{2}(E_0e^{i(k(\omega)z-\omega t)}+c.c)$, while its
phase is $\varphi=k(\omega)z-\omega t$. We assume this phase of
the field constant during the propagation through the medium. The
group index obtained is written by
\begin{equation}
N_g=n_r(\omega)+\omega\frac{\partial
n_r(\omega)}{\partial\omega},
\end{equation}
The corresponding dispersion of the group velocity of the light
pulse is:
\begin{equation}
D_{v_g}=\frac{\partial
}{\partial\omega}(v^{-1}_g)=\frac{1}{c}[2\frac{\partial
n_r(\omega)}{\partial\omega}+\omega\frac{\partial
n^2_r(\omega)}{\partial\omega^2}].
\end{equation}
Furthermore, the complex wave-number $k(\omega)$ can be expanded
via Taylor series in terms of group index as:
\begin{eqnarray}
k(\omega)&=&\frac{N^{(0)}_g\omega}{c}+\frac{1}{2!}\left( \omega -\omega _{0}\right) ^{2}\frac{1}{c}\frac{\partial N_{g}}{\partial \omega
}|_{\omega\longrightarrow\omega_0}\nonumber\\&&+\frac{1}{3!}\left( \omega -\omega _{0}\right) ^{3}\frac{1}{c}\frac{\partial^2 N_{g}}{\partial \omega^2
}|_{\omega\longrightarrow\omega_0}\nonumber\\&&+\frac{1}{4!}\left( \omega -\omega _{0}\right) ^{4}\frac{1}{c}\frac{\partial^3 N_{g}}{\partial \omega^3
}|_{\omega\longrightarrow\omega_0}....
\end{eqnarray}
where $N^{(0)}_g$, is the group index of a medium at the central
frequency $\omega_0$. The transit time of the pulse through the
medium is $T=N_g L/c=k_1 L$, and $\Delta T=[L\partial
N_g/c\partial\omega]\Delta w$ represents the spread in transient
time, for $\Delta w$ being the frequency bandwidth of the pulse.
Under the condition $\Delta T<\tau _{0}$, there is no significant
distortion in the system. We denote $\tau _{0}$ for the
characteristic pulse width. The distortion in the pulse
propagating through the proposed medium becomes negligible,
therefore $\partial N_g/\partial \omega=0$. Generally, it is the
most favorable condition for an experiment while having the
advantage of extremum of negative $N_g$. In this connection, our
proposed system is based on phenomenon which has lossless
anomalous dispersion regions for the probe pulse propagation to
have minimum contribution from the higher order terms in
$k(\omega)$. Satisfactorily, under these conditions the pulse
shape at the output remains almost unchange while it modify the
advance time $\tau_d=\frac{L(N_g-1)}{c}$, significantly. To
developed the formalism and to study the nature of the pulse at
the output we incorporate the terminology as $n_r$ and $N_g$ being
the refractive index and the group index, respectively.
Furthermore, $\partial n_r/\partial\omega$ is related to
dispersion and $\partial N_g/\partial \omega$ describe the
distortion. Obviously, the real and imaginary parts of the higher
order terms of $k_j(\omega)$ are related to the dispersion and
phase distortion, and to the gain (absorption) and amplitude
distortion, respectively. The output pulse $S_{out}(\omega)$,
after the propagating through the medium can be related to the
input pulse $S_{in}(\omega)$ with the transfer function
$H(\omega)$, being a convolution. The transfer function for our
system is written as $S_{out}(\omega)=H(\omega)S_{in}(\omega)$,
where $H(\omega)=e^{-ik(\omega)L}$ is the transfer function for
our system. Now, we choose a Gaussian input pulse of the form
\begin{eqnarray}
S_{in}(t)=\exp[-t^2/\tau^2_0]\exp[i(\omega_0+\xi)t],
\end{eqnarray}
for $\xi$, being the upshifted frequency of the empty cavity. The
Fourier transforms of this function is then written by
$S_{in}(\omega)=\frac{1}{\sqrt{2\pi}}\int^{\infty}_{-\infty}S_{in}(t)e^{i\omega
t}d t$. The expression for the input signal is calculated as
\begin{eqnarray}
S_{in}\left( \omega \right)  &=&\tau _{o}/\sqrt{2}\exp \left[
-\left( \omega -\omega _{o}-\xi \right) ^{2}\tau^2 _{o}/4\right],
\end{eqnarray}
By virtue of the convolution theorem the output $S_{out}(t)$, can
be related to the input pulse via the transfer function using
inverse Fourier transforms as
$S_{out}(t)=\frac{1}{\sqrt{2\pi}}\int^{\infty}_{-\infty}S_{int}(\omega)H(\omega)e^{i\omega
t}d\omega$. Evidently, the response of the advance time is
significant with the Doppler broadening and we have drastic
enhancement in the superluminality of the Gaussian light pulse. In
this connection, it is worthwhile to analyze the output pulse
shape distortion for high enough perturbation limit. To the fourth
order of the group index it is given by

\begin{eqnarray}
S_{out}(t)&=&\frac{\tau_0\sqrt{c}}{\sqrt{2i
n_1L+c\tau^2_0}}[1-\frac{i(6F_1F_2+F^3_2)n_2
L}{48cF^3_1}]\nonumber\\&&\times\exp[-\frac{\xi^2\tau^2_0}{4}+i(t-\frac{n_0
L}{c})\omega_0]\nonumber\\&&\times\exp[F_2+\frac{cF^2_2}{(2i
n_1L+c\tau^2_0)}]
\end{eqnarray}
\begin{eqnarray}
F_1=\frac{2i n_1L+c\tau^2_0}{4c}
\end{eqnarray}
\begin{eqnarray}
F_2=\frac{c\xi\tau^2_0-2i n_0L+2ict}{2c}
\end{eqnarray}
Where  $n_1=\partial N_g/\partial\omega$ and $n_2=\partial^2
N_g/\partial\omega^2$. The nature of the output pulse after
propagating thought the medium can be analyzed from the above
analytical expression. Since there are multiple anomalous regions
in the dispersion region therefore we provide the quantitative and
graphical analysis for one of these regions for the pulse
distortion measurement in the next section.
\section{Results and Discussion}
\begin{figure}[t]
\centering
\includegraphics[width=2.5in]{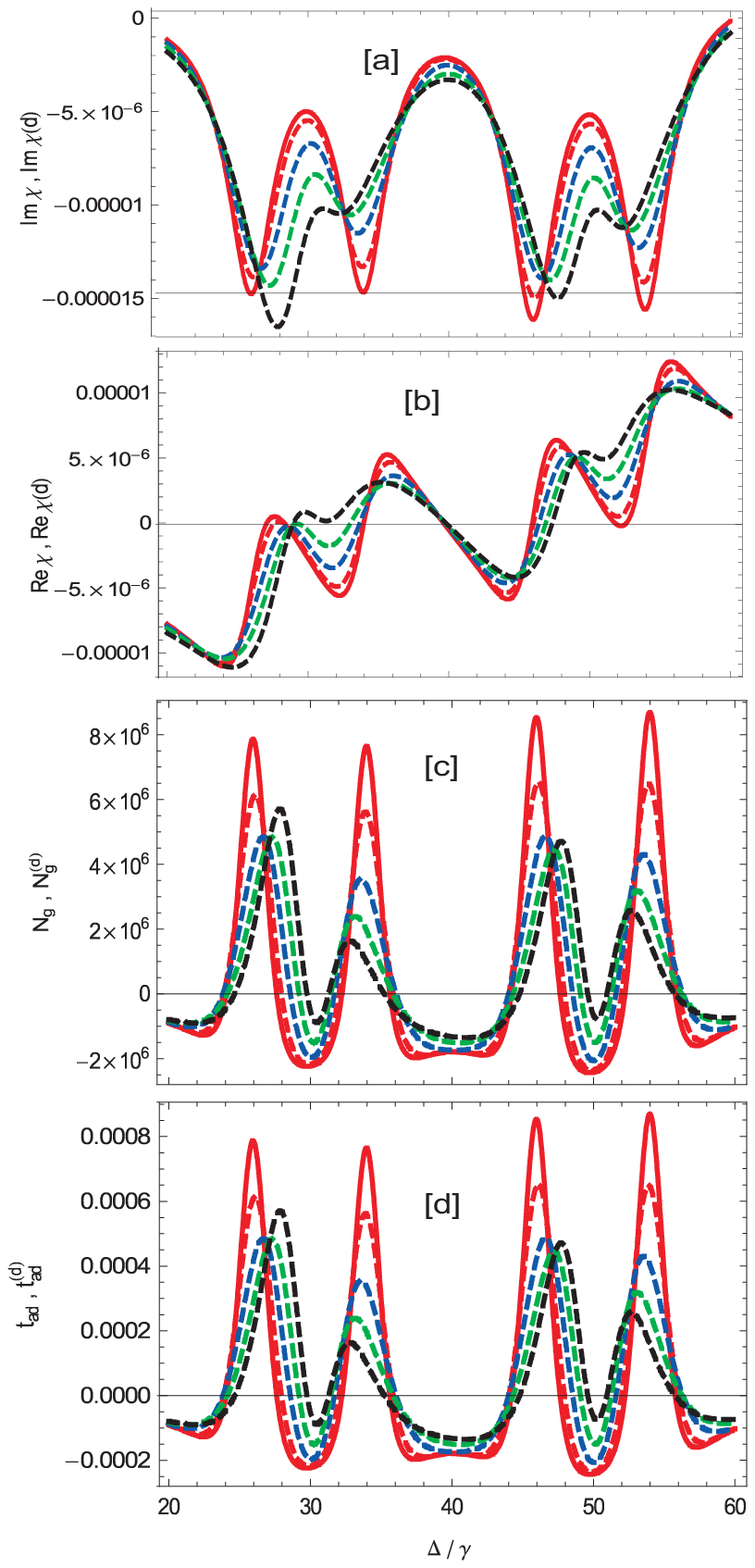}
\caption{Dispersion, gain, group index and group delay time
against $\frac{\Delta}{\protect
\gamma}$ such that $\protect\gamma%
=1MHz$, $\Gamma_{ad}=\Gamma_{ac}=\Gamma_{ab}=
\Gamma_{cd}=\Gamma_{bd}=2.01\protect\gamma$, $\protect
\delta_3=0\protect\gamma$, $\protect\lambda=586.9nm$,
$\protect\omega_{ac}=10^3 \protect\gamma$, $
\protect\delta_1=30\protect\gamma$,
$\protect\delta_2=50\protect\gamma$, $ \Omega_1=4\protect\gamma$,
$\Omega_2=7\protect\gamma$, $\Omega_3=8\protect \gamma$,
$V_D=0\protect\gamma$(solid red), $V_D=2\protect\gamma$(dashed
red), $V_D=4\protect\gamma$(blue dashed),
$V_D=6\protect\gamma$(green dashed), $V_D=8\protect\gamma$(black
dashed).When the intensity of control field is larger then the
pump field. The gain doublet are present and the group index are
decrease both in positive/negative domain with the increase of
doppler width.  The superlominal propagation is reduced with
doppler width  at two photons resonance pionts
$\Delta=\delta_{i=1,2}=30\gamma,50\gamma$ and between the two pair
of the gain doublet region, ($\Delta=40\gamma$).} \label{figure1}
\end{figure}
\begin{figure}[t]
\centering
\includegraphics[width=2.8in]{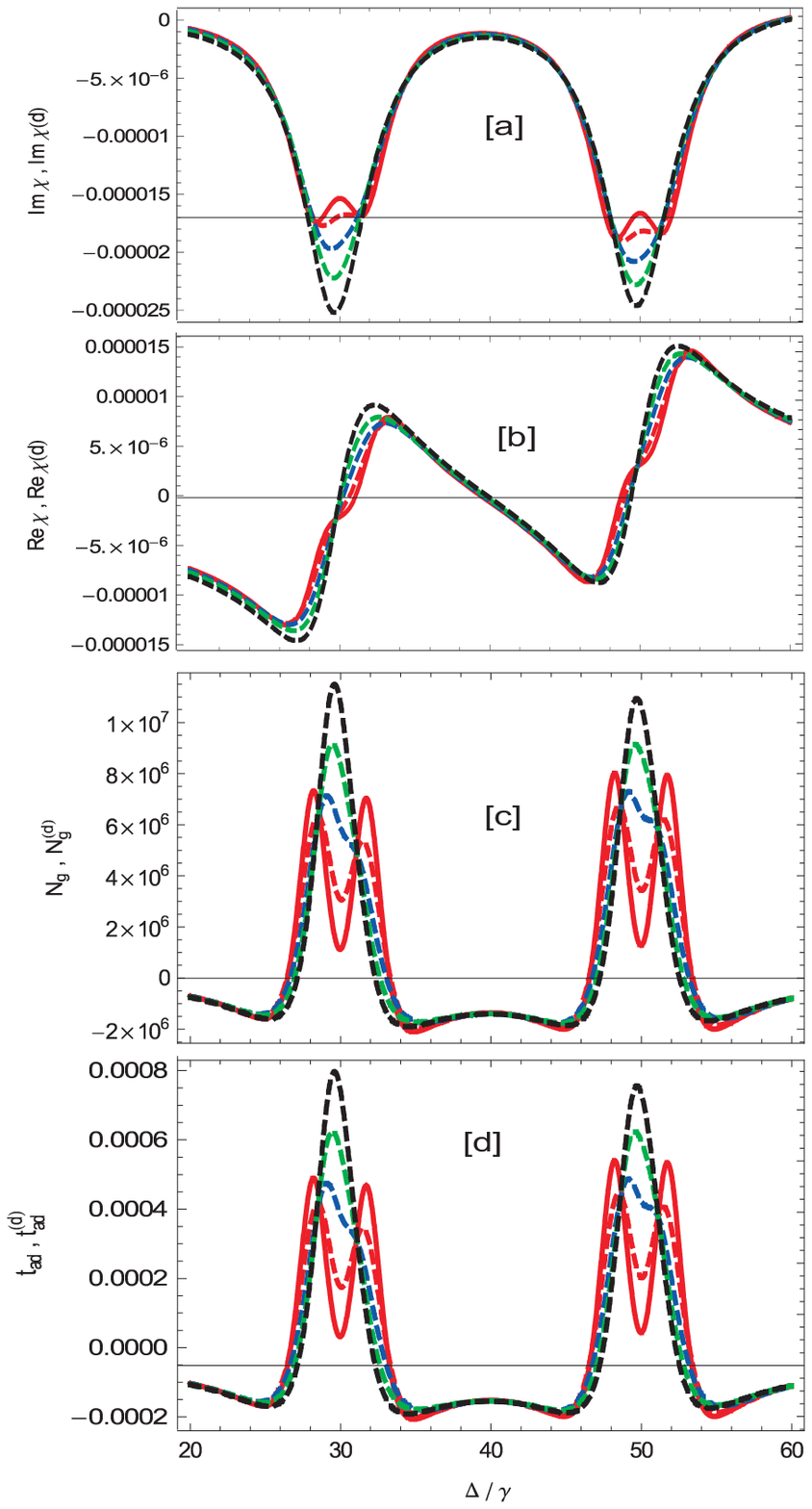}
\caption{Dispersion, gain, group index and group delay time
against $\frac{\Delta}{\protect
\gamma}$ such that $\protect\gamma%
=1MHz$, $\Gamma_{ad}=\Gamma_{ac}=\Gamma_{ab}=
\Gamma_{cd}=\Gamma_{bd}=2.01\protect\gamma$, $\protect
\delta_3=0\protect\gamma$, $\protect\lambda=586.9nm$,
$\protect\omega_{ac}=10^3 \protect\gamma$, $
\protect\delta_1=30\protect\gamma$,
$\protect\delta_2=50\protect\gamma$, $ \Omega_1=4\protect\gamma$,
$\Omega_2=7\protect\gamma$, $\Omega_3=3.5\protect \gamma$,
$V_D=0\protect\gamma$(solid red), $V_D=2\protect\gamma$(dashed
red), $V_D=4\protect\gamma$(blue dashed),
$V_D=6\protect\gamma$(green dashed), $V_D=8\protect\gamma$(black
dashed). When the intensity of control field is smaller then the
pump fields.The gain doublet is vanish with the increase of
doppler width. The sublominal propagation is suddenly enhance at
two photons resonance pionts
$\Delta=\delta_{i=1,2}=30\gamma,50\gamma$.} \label{figure1}
\end{figure}
\begin{figure}[t]
\centering
\includegraphics[width=2.5in]{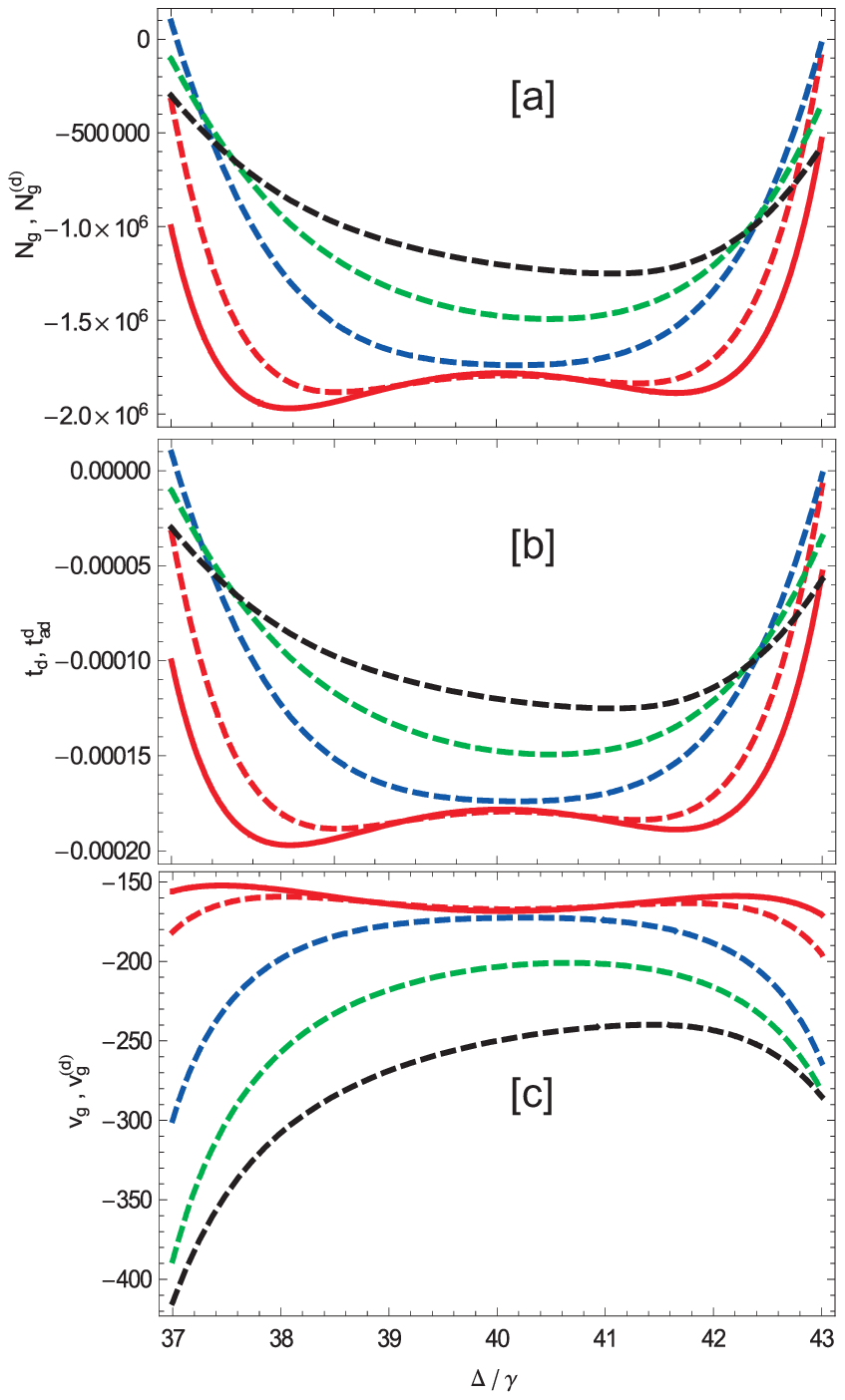}
\caption{Group index group advance time and group velocity versus
$\frac{\Delta}{\protect\gamma}$ such that $\protect\gamma%
=1MHz$, $\Gamma_{ad}=\Gamma_{ac}=\Gamma_{ab}=
\Gamma_{cd}=\Gamma_{bd}=2.01\protect\gamma$, $\protect
\delta_3=0\protect\gamma$, $\protect\lambda=586.9nm$,
$\protect\omega_{ac}=10^3 \protect\gamma$, $
\protect\delta_1=30\protect\gamma$,
$\protect\delta_2=50\protect\gamma$, $ \Omega_1=4\protect\gamma$,
$\Omega_2=7\protect\gamma$, $\Omega_3=8\protect \gamma$,
$V_D=0\protect\gamma$(solid red), $V_D=2\protect\gamma$(dashed
red), $V_D=4\protect\gamma$(blue dashed),
$V_D=6\protect\gamma$(green dashed), $V_D=8\protect\gamma$(black
dashed). The group index is reduce in negative domain with the
increase of doppler width. The advance time is also reduce with
the doppler width, but the group velocity is increase in negative
domain. More negative group index is more advance time and smaller
negative group velocity. Larger superlominality occur at more
negative group index. The Fig[a,b,c] clearly show that the
superlominality is reduced with the doppler effect.}
\label{figure1}
\end{figure}
\begin{figure}[t]
\centering
\includegraphics[width=3in]{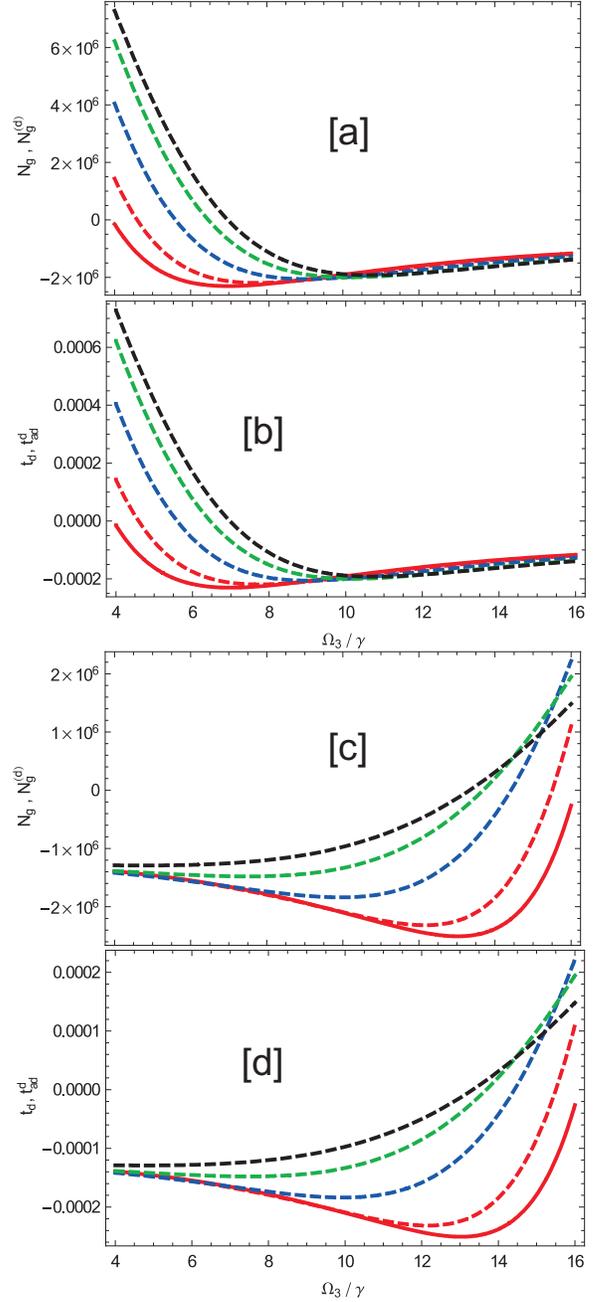}
\caption{Group index and Group delay/advanced time versus
$\frac{\Omega_3}{\protect\gamma}$ such that $\protect\gamma%
=1MHz$, $\Gamma_{ad}=\Gamma_{ac}=\Gamma_{ab}=
\Gamma_{cd}=\Gamma_{bd}=2.01\protect\gamma$, $\protect
\delta_3=0\protect\gamma$, $\protect\lambda=586.9nm$,
$\protect\omega_{ac}=10^3 \protect\gamma$, $
\protect\delta_1=30\protect\gamma$,
$\protect\delta_2=50\protect\gamma$, $ \Omega_1=4\protect\gamma$,
$\Omega_2=7\protect\gamma$,[a,b]$\Delta=30\gamma$[c,d]$\Delta=40\gamma$,
$V_D=0\protect\gamma$(solid red), $V_D=2\protect\gamma$(dashed
red), $V_D=4\protect\gamma$(blue dashed),
$V_D=6\protect\gamma$(green dashed), $V_D=8\protect\gamma$(black
dashed). } \label{figure1}
\end{figure}
\begin{figure}[t]
\centering
\includegraphics[width=2.8in]{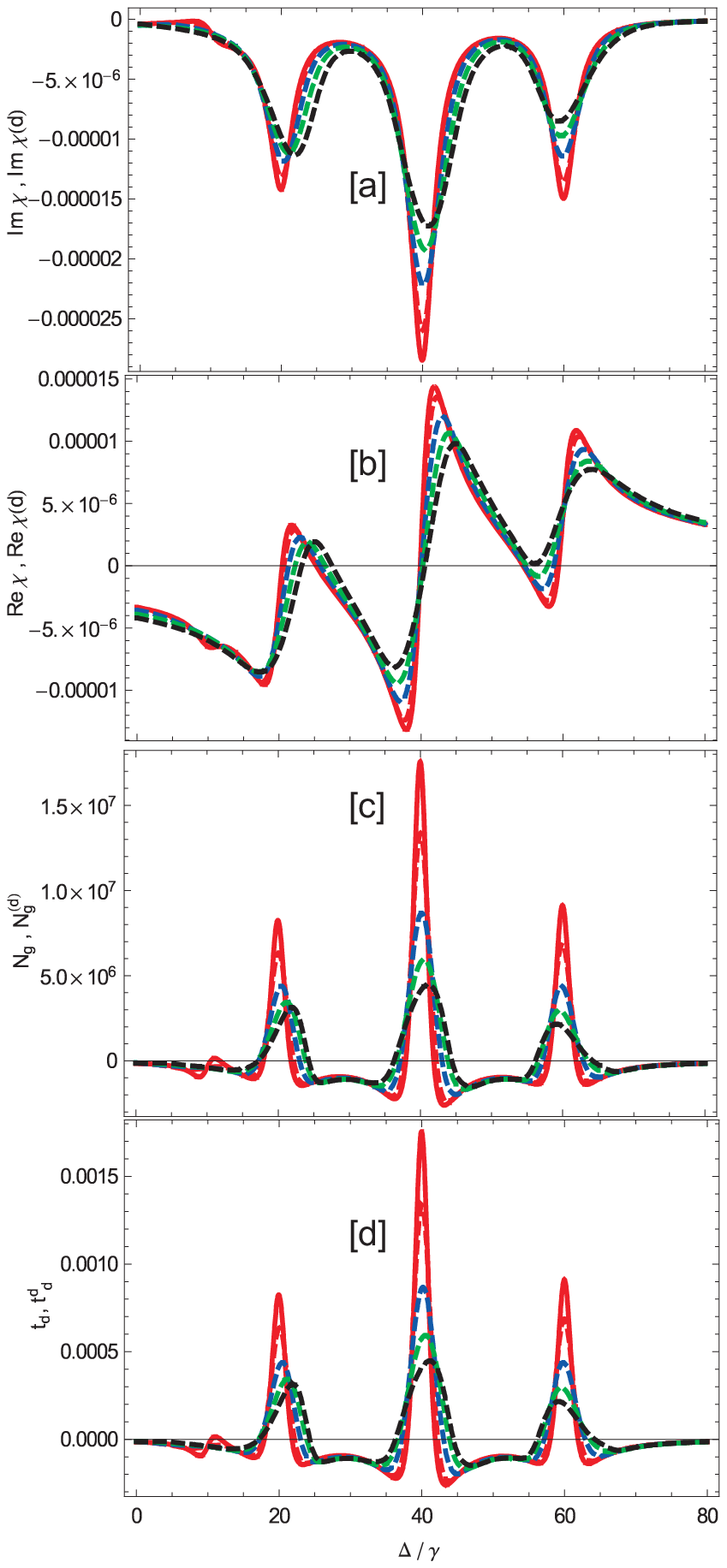}
\caption{Dispersion, gain, group index and group delay time
against $\frac{\Delta}{\protect
\gamma}$ such that $\protect\gamma%
=1MHz$, $\Gamma_{ad}=\Gamma_{ac}=\Gamma_{ab}=
\Gamma_{cd}=\Gamma_{bd}=2.01\protect\gamma$, $\protect
\delta_3=0\protect\gamma$, $\protect\lambda=586.9nm$,
$\protect\omega_{ac}=10^3 \protect\gamma$, $
\protect\delta_1=30\protect\gamma$,
$\protect\delta_2=50\protect\gamma$, $ \Omega_1=4\protect\gamma$,
$\Omega_2=7\protect\gamma$, $\Omega_3=20\protect \gamma$,
$V_D=0\protect\gamma$(solid red), $V_D=2\protect\gamma$(dashed
red), $V_D=4\protect\gamma$(blue dashed),
$V_D=6\protect\gamma$(green dashed), $V_D=8\protect\gamma$(black
dashed).} \label{figure1}
\end{figure}
\begin{figure}[t]
\centering
\includegraphics[width=3in]{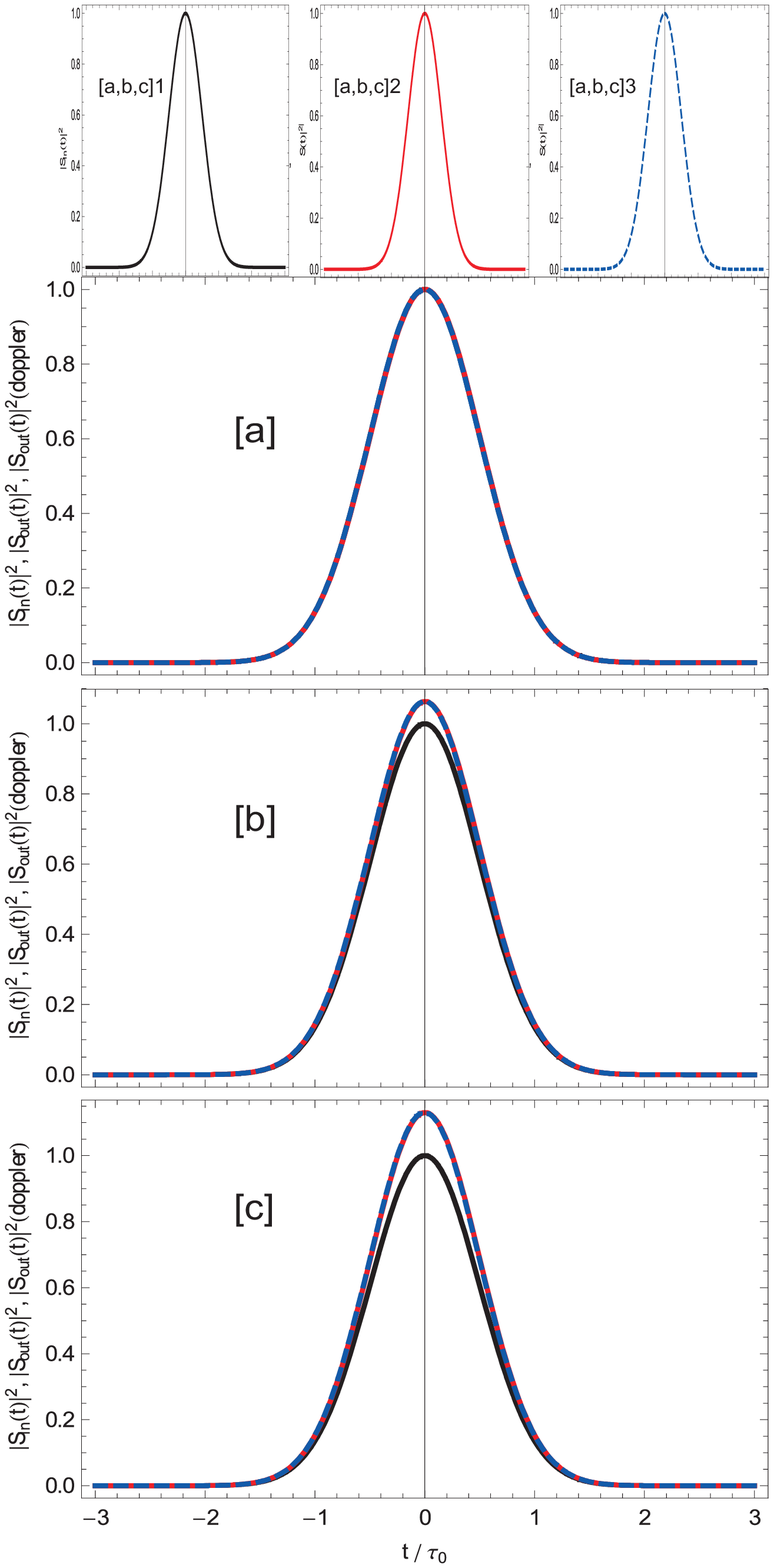}
\caption{The intensity of a Gaussian pulse at input and output Vs
the normalized time for both the increasing and decreasing
function of the group index [a] $\xi=0 MHz $[b]$\xi=0.005 MHz $,
[c] $\xi=0.01 MHz $, $\tau_0=3.50\mu s$, $c=3\times10^{8}m/s$,
$L=0.03m$, $\omega_0=970 \gamma$,
$\Gamma_{ad}=\Gamma_{ac}=\Gamma_{ab}=\Gamma_{cd}=2.01\gamma$,
$\Gamma_{bd}=0.01\gamma$, $\delta_3=0\gamma$, $\lambda=586.9nm$,
$\omega_{ac}=10^3\gamma$, $V_D=4\gamma$, $\delta_1=30\gamma$,
$\delta_2=50\gamma$, $\Omega_1=4\gamma$, $\Omega_2=7\gamma$,
$\Omega_3=8\gamma$} \label{figure1}
\end{figure}
We proceed with inspection of our main results presented in the
Eqs. (4,7,8,9,10,14,16) for dispersion gain group index time delay
and pulse shape distorsion, when there is no Doppler broadening
effect in the system and for the system when there is its maximum
effect. To investigate the soundness of the results of our
proposed system, we kept checks on the main results under some
approximations. First, we turn off the control field in our system
which leads to the famous scheme of Wang \emph{et al}. In this
approximation, the results of our system are in complete agreement
with the said reference under the same set of parameters and we
have two closely spaced gain-peak behavior. The reduction of the
analytical results to the Wang \emph{et al}. scheme is
satisfactory, representing the soundness of our results. Moreover,
in this approximated scheme we further exclude one of two coherent
pump fields, then the atomic system reduces to a single Raman gain
scheme where only one gain peak exists. Further, in this set-up if
we include the control field with a fourth level then this forms
N-type scheme with one coherent pump field similar to the one
discussed by Agarwal \textit{et al.} \cite{TNGS2007}. The behavior
of the gain profile of the single Raman gain scheme is modified in
this N-type case. Here, the single gain feature is modified by the
control field into two closely spaced gain-peak behavior unlike
the single Raman gain scheme. The splitting of the single gain
peak into two is the manifestation of the dressing of the ground
state in a doublet. All these facts also agree with the results
presented in Ref. \cite{TNGS2007}.

Next, we focus on the main results of our atomic system and are
committed (1) to discuss the underlined changes in the physical
behavior of this system as compared to the earlier related
literatures (2) to discuss the subluminal and superluminal
behavior of the light pulse propagating through their associated
dispersion regions, (3) to discuss the induced Doppler broadening
effect in the system and (4) to discuss that how the control field
resolves the fundamental issue of the Wang \emph{et al.} scheme
where the singularities and other effects prevent the system to be
Doppler broadened. This is a very important aspect of the
experiment which cannot be incorporated due to singularities
appeared at the denominator of the two terms of susceptibility.
 This fact limits the usefulness of
their experiment. However, in present system this issue is
resolved due to the nonsingular terms appeared at the denominator
of the two terms of the susceptibility.

In the proposed scheme, the Doppler effect is maximum for our
considered appropriate orientation of the counter propagating
driving fields relative to the atomics motion. The Doppler width
of our system depends linearly on temperature of the cavity and
density of the medium. Then, greater the temperature of the cavity
or greater the density of the medium or both, the larger is the
Doppler width. Subsequently, the smaller the negative group index,
the manimum is the superluminal effect on the propagating Gaussian
pulse. As a result, the group velocity of the superluminal
Gaussian pulse can be reduce to its highest ultimate limit for
this system.

The real and imaginary parts of the susceptibilities correspond to
the gain and dispersion as well as corresponding group index and
time delaty/advance are plotted against the probe field detuning
for the two cases presented in the Eqs. (4),(7)-(8,9,10) for the
same set of parameters and when $\Delta=\delta_1 $ or
$\Delta=\delta_2$ [see Fig. 2]. The spectral profile of the gain
of the system shows two pairs of the gain doublet. This behavior
is completely different from the gain assisted system of Wang
\emph{et al.}, where only one pair of a doublet exists. The change
in the scenarios of the gain profile for this system is important
as it provides multiple anomalous regions in the dispersion medium
for the superluminal pulse propagation. This kind of regions can
be found in the recently reported works in the context of EIT
\cite{EIT}. Further, the Doppler broadening effect of the system
reduced the superluminal profile drastically. Furthermore, the
control field can be used to control the anomalous regions very
efficiently. The physics of these under lying changes is very
interesting. The coherently driven control field splits one of the
ground energy levels in a doublet. In this case the photons
associated with the two coherent pump fields are added to the
probe field through two paths created by the control field in the
form of a two dressed-state. Thus, there are two pairs of the gain
doublet. The smaller (larger) the strength of the control field,
the lesser (greater) is the space between the two dressed states.
Consequently, the smaller (larger) is region between the two gain
doublets. Evidently, this leads to the control of the anomalous
dispersion regions if and only if the detuning of the control
field is small. However, when this detuning becomes large,
splitting of the ground state disappears. As a result, two pairs
of the doublet-gain reduce to one pair and exhibits the double
Raman gain processes like the Wang \emph{et al}. This behavior
agrees to what we expect for large detuning where the control
field interacts no more with the upper added fourth level.
Consequently, two dressed ground states associated with the real
ground energy level disappear. Further, the imaginary part of
$\chi(d)$, shows a typical gain dip over the imaginary part of
$\chi$ and the doublet is disapear with the increase of doppler
width[see Fig. 3]. The slope of dispersion in these gain dip
regions are normal for both $\chi^(d)$ when the strenght of
control field is smaller then the pump field, which shows slow
light propagation. The group indices in these dispersion regions
are positive and enhances with the Doppler-broadened system.
Interestingly, the gain increase significantly with the Doppler
broadening effect between the two gain peaks around the two
photons resonance condition i.e., $\Delta=\delta_1$ or
$\Delta=\delta_2$, there is anomlous region for both the
$\chi^(d)$, and $\chi$. Furthermore, the anomalous dispersion
associated with $\chi^(d)$, is less pronounced and steeper than
the dispersion of $\chi$. The group index close to the resonance
condition i.e., $\Delta=30\gamma,\Delta=50\gamma$, in the
anomalous dispersion region reduces for the Doppler-broadened
system and we have $N_g=-2.2284\times10^6$ and
$N_g(d)=-1.09624\times10^6$. Their corresponding group velocities
are given by $v_g=-c/2.2284\times10^6=-134.626m/s$ and $v^d_g=
-c/1.09624\times10^6=-273.664m/s$, respectively. The negative
group advance time for $v_g=-c/ 2.2284\times10^6$, is $-22.28\mu
s$, and for $v^d_g=-c/1\times10^8$, is $-10.96\mu s$, across the
$3 cm$, length of the dispersion medium if the doppler width is
$V_D=2\gamma$. The pulse at $v^d_g$, is slow down than the pulse
of $v_g$. The reduce advance time is $ -11.32\mu s$[see Fig.
2[c,d]] further increase of doppler width reduce the advance time
repidly. The plots traced for group index group advance time and
group velocity against probe detuning as a increasing function of
doppler width around the point $\Delta=40\gamma $ are also shown
in Fig .4. The group index is reduce in negative domain when the
doppler width is increase stepwise from
$0\gamma,2\gamma,4\gamma,6\gamma,8\gamma$ . The group index varies
stepwise from
$N^{d}_g=-1.78\times10^6,-1.79\times10^6,-1.73\times10^6,-1.47\times10^6,-1.20\times10^6$
while the group velocity varies from
$v^d_g=[-168.29,-167.20,-172.62,-203.42,-249.95]m/s$, and the
group advance time varies from $t^d_{ad}=[-17.82 ,-17.94,-17.37\mu
s,-14.74,-12.002]\mu s$ as shown in Fig4.[a,b,c]. Fig5 show
variation of group index with control field at different values of
doppler widths. At high intensity of the control field
$\Omega_3=20\gamma$, the region between the two pair of gain
doublet is disapear and simply three gain dipth are observe. The
gain dipth is reduce with the increase of doppler width. The slope
of dispersion in these gain depth region are normal and reverse
manipulated with the doppler width. The group index in these
regions are positive and reduce doppler width,which enhance the
group velocity in positive domain. There is large time delay in
the abscent of doppler effect. The delay time is reduce with
doppler effect Fig6.[a,b,c,d].

Furthermore, it is noteworthy that a larger negative group index
is not a necessary and sufficient condition for practical
application. It is necessary to choose a system with minimum
possible losses (gain) along with its best negative group index
characteristics. The scheme of Wang \emph{et al.} due to its very
less gain at the anomalous region of the medium is responsible for
its successful demonstration in a laboratory. Therefore, a scheme
having significant gain at anomalous region may distort the output
pulse and may lead to unpractical condition. Obviously, the
response of the advance time is significant with the Doppler
broadening effect and we have the drastic enhancement in the
superluminality of the pulse. In this connection, it is worthwhile
to analyze the output pulse shape distortion for high enough
perturbation limit. The analytical results are provided in the
text to the fourth order of the group index for the output
Gaussian pulse. The Gaussian pulse shape is presented in the time
domain by transforming the angular-frequency dependent input
Gaussian pulse using transfer function of the medium as a
convolution. In this way we provide the output pulse in time
domain to compare with the input Gaussian pulse and almost no
distortion is seen. This behavior confirms the characteristics
similar to the Wang et al. model for Doppler-free and
Doppler-broadened systems and obeys an ideal behavior.
Consequently, the imbedded Doppler effect which is responsible for
the drastic increase of the group index is still practical. The
anomalous regions are always there in the mid of each pair of the
gain lines with the providence of good quality characteristics for
the pulse shape at the output of the medium. Quantitatively, we
choose one of the lossless anomalous regions appears at
$\Delta=30\gamma$ and the atomic transition frequency as
$\omega_{ac}=1000\gamma$. Furthermore, we also consider the
central frequency of the probe field at $\Delta=30\gamma$, as
$\omega_0=970\gamma$. The detuning of the probe field is written
by $\Delta=\omega_{ac}-\omega$, where $\omega=2\pi\nu_p$. The
values of group indices for the two cases in the lossless region
are $n_0 =-2.22\times 10^{6}, $$ n_{0d}=-1.95\times 10^{6}$.
However, the first order derivatives of group indices are
calculated as $n_1=2.65\times10^7$, $n_{1d}=-3.46\times10^7$ and ,
the second order derivatives of group indices are
$n_2=1.26\times10^{9}$, $n_{2d}=-4.54\times10^8$.
 In these cases the real parts correspond to the
amplitude distortion while the imaginary parts correspond to the
phase distortion of the propagating Gaussian pulse through our
proposed medium.

Traditionally, the physical interpretation of superluminality by
virtual reshaping is now not reasonable \cite{Wangpra,book1,book2}
with the explanation of amplification of the front edge with the
relatively absorbtion of its tail. In the Wang \emph{et al.}
experiment the band-width of the probe pulse is chosen very
smaller than the separation of the two gain peaks. In the pulse
distortion measurement we kept in our system the probe pulse
bandwidth much smaller than gain lines separation to avoid
resonances with the Raman transitions frequencies for the probe
pulse. Consequently, there is no amplification of the front edge
of the pulse. Moreover, the average time of stay of an atom inside
the volume of the Raman probe beam is also shorter than than FWHM
of the probe pulse. This means that both the front edge and the
tail would be amplified if the atoms of the medium amplify the
probe pulse. However, this is not in accord with the earlier
claims and even with our displayed results. Obviously, the
superluminal light propagation arises due to the anomalous
dispersion region created by the two nearby Raman gain resonances.
In fact, if the gain becomes large, its effect appears as a
compression of the pulse \cite{Wangpra}. This fact is very clearly
observable from graphical analysis of our pulse distortion
measurement as shown in the Fig. 7. The detail analysis reveals
that the pulse distortion measurement fully agree with the Wang
\emph{et al.} studies even with the fourth order perturbation
limit. Evidently, in our system it is shown that pulse shape is
preserved over the very small region in between the gain lines for
some specific value of the upshift frequency. The shifting of this
upshift frequency toward the either gain line results in
compression or enhancing the pulse shape of the input probe field
corresponding to an increase or decrease in the gain,
respectively. In fact it is the consideration of probe pulse-width
much smaller than the gain lines separation which results in the
transparency over a very narrow domain of the upshifting
frequency. The movement of the shifting frequency to either sides
of the gain line results in the suppression and enhancement of the
shape of the probe pulse and are the essence of the pulse
distortion as shown in Fig. 7. Unlikely, this behavior does not
agree with the earlier interpretation and is more likely with Wang
et al, presented results and interpretation. Satisfactorily, we
are providing analytical results to the literature where this
behavior can be predicted and proceeded beyond the literature.
\section{Summary}
We proposed an N-shaped 4-level atomic system driven by two pump
fields, a control and a probe field appropriately. The proposed
scheme displays interesting and novel results of the two pairs
Raman gain processes having various advantages than the double
Raman gain scheme of Wang's \emph{et al.} Generally, the present
system consists of two controllable pairs of double Raman gain
peaks with control of the control field. However, the Doppler
broadening effect induced by the control field in the system
enhances significant enhancement in the superluminality of the
probe Gaussian light pulse. Consequently, under experimentally
feasible parameters the advanced time is shorter by $76.12 ms$
than the advance time of Wang \emph{et al} Doppler insensitive
experiment. The proposed two-paired gain system is lossless with
the extra advantages of Doppler sensitivity and multiple
controllable lossless anomalous regions while having almost
undistorted output pulse. In fact, making an object hidden in
space (time) from a physical observer
\cite{Livale2009,Gab2009,Leonh2009,chen2007,Mc2011}, is a
laboratory reality which requires negative group indices
\cite{Wang} with almost undistorted pulse shape at the output of
the medium. Similarly, superluminal light pulse can also be used
for imaging \cite{RYANT2012}. However, due to limitation of the
current technology, it is appealing to explore mechanisms to
create the hole for cloaking to microsecond and millisecond and to
improve a best quality image measurement \cite
{Leonh2009,Fridman2012,Mc2011}. In this connection, the present
scheme may provide this ground to get improve the applied aspects
of the superluminality and is easily adjustable with the current
technology.
\section{Appendix-A}
The dynamical equations obtained from Eq. (2) for slowing varying
amplitudes are given by
\begin{eqnarray}
\overset{\cdot }{\overset{\sim }{\rho
}}_{aa}&=&\frac{i}{2}[(\Omega_1+\Omega_2)\cos\phi+i(\Omega_1-\Omega_2)\sin\phi]\widetilde{\rho}_{da}\nonumber\\&&-\frac{i}{2}[(\Omega^*_1+\Omega^*_2)\cos(\phi)-i(\Omega^*_1-\Omega^*_2)\sin(\phi)]\widetilde{\rho}_{ad}\nonumber\\&&+\frac{i}{2}\Omega_p\widetilde{\rho}_{ca}-\frac{i}{2}\Omega^*_p\widetilde{\rho}_{ac}-(\Gamma_{ad}+\Gamma_{ac})\widetilde{\rho}_{aa},
\end{eqnarray}
\begin{eqnarray}
\overset{\cdot }{\overset{\sim }{\rho
}}_{bb}=-(\Gamma_{bc}+\Gamma_{bd})\widetilde{\rho}_{bb}+\frac{i}{2}
( \Omega_3\widetilde{\rho}_{cb} - \Omega^*_3\widetilde{\rho}_{bc},
\end{eqnarray}
\begin{eqnarray}
\overset{\cdot }{\overset{\sim }{\rho
}}_{cc}=\frac{i}{2}\Omega^*_3\widetilde{\rho}_{bc}-\frac{i}{2}\Omega_3\widetilde{\rho}_{cb}+\frac{i}{2}\Omega^*_p(\widetilde{\rho}_{ac}-\widetilde{\rho}_{ca}),
\end{eqnarray}
\begin{eqnarray}
\overset{\cdot }{\overset{\sim }{\rho
}}_{bc}=(i\delta_3-\Gamma_{bc})\widetilde{\rho}_{bc}+\frac{i}{2}
 \Omega_3(\widetilde{\rho}_{cc} - \widetilde{\rho}_{bb})-\frac{i}{2}\Omega_p\widetilde{\rho}_{ba},
\end{eqnarray}
\begin{eqnarray}
\overset{\cdot }{\overset{\sim }{\rho
}}_{dd}&=&\frac{i}{2}[(\Omega^*_1+\Omega^*_2)\cos\phi-i(\Omega^*_1-\Omega^*_2)\sin\phi]\widetilde{\rho}_{ad}\nonumber\\&&-\frac{i}{2}[(\Omega_1+\Omega_2)\cos\phi+i(\Omega_1-\Omega_2)\sin\phi]\widetilde{\rho}_{da}\nonumber\\&&+\Gamma_{ad}\widetilde{\rho}_{aa}+\Gamma_{bd}\widetilde{\rho}_{ab},
\end{eqnarray}
\begin{eqnarray}
\overset{\cdot }{\overset{\sim }{\rho
}}_{ad}&=&\frac{i}{2}[(\Omega_1+\Omega_2)\cos\phi+i(\Omega_1-\Omega_2)\sin\phi](\widetilde{\rho}_{dd}-\widetilde{\rho}_{aa})\nonumber\\&&+
\left[
\frac{i}{2}\left( \delta _{1}+\delta _{2}\right) -\Gamma
_{ad}\right] \overset{\sim }{\rho
}_{ad}+\frac{i}{2}\Omega_p\widetilde{\rho}_{cd},
\end{eqnarray}
\begin{eqnarray}
\overset{\cdot }{\overset{\sim }{\rho
}}_{ac}&=&(i\Delta-\Gamma_{ac})\widetilde{\rho}_{ac}+\frac{i}{2}
 \Omega_p(\widetilde{\rho}_{cc}-\widetilde{\rho}_{aa})-\frac{i}{2}\Omega_3\widetilde{\rho}_{ab}\nonumber\\&&+\frac{i}{2}[(\Omega_1+\Omega_2)\cos\phi\nonumber\\&&+i(\Omega_1-\Omega_2)\sin\phi]\widetilde{\rho}_{dc},
\end{eqnarray}
\begin{eqnarray}
\overset{\cdot }{\overset{\sim }{\rho
}}_{dc}&=&\frac{i}{2}[(\Omega^*_1+\Omega^*_2)\cos\phi-i(\Omega^*_1-\Omega^*_2)\sin\phi]\widetilde{\rho}_{ac}\nonumber\\&&-\frac{i}{2}\Omega_3\widetilde{\rho}_{db}+\frac{1}{2}\left[
i\left( 2\Delta -\delta _{1}-\delta _{2}\right) -2\Gamma
_{cd}\right] \overset{\sim }{\rho
}_{dc}\nonumber\\&&-\frac{i}{2}\Omega_p \widetilde{\rho}_{da} ,
\end{eqnarray}
\begin{eqnarray}
\overset{\cdot }{\overset{\sim }{\rho
}}_{ab}&=&\frac{i}{2}[(\Omega_1+\Omega_2)\cos\phi+i(\Omega_1-\Omega_2)\sin\phi]\widetilde{\rho}_{db}\nonumber\\&&+
\frac{i}{2}\Omega_p\widetilde{\rho}_{cb}-\frac{i}{2}\Omega^*_3\widetilde{\rho}_{ac}+(i(\Delta-\delta_3)-\Gamma_{ab}))\widetilde{\rho}_{ab},
\end{eqnarray}
\begin{eqnarray}
\overset{\cdot }{\overset{\sim }{\rho }}_{bd}&=&\frac{1}{2}\left[
i\left( 2\delta _{3}+\delta _{1}+\delta _{2}-2\Delta \right)
-2\Gamma _{bd}\right] \overset{\sim }{\rho }_{bd}\nonumber\\&& -
\frac{i}{2}[(\Omega_1+\Omega_2)\cos\phi+i(\Omega_1-\Omega_2)\sin\phi]\widetilde{\rho}_{ba}\nonumber\\&&+\frac{i}{2}
 \Omega_3\widetilde{\rho}_{cd},
\end{eqnarray}
In the above equations the phase $\phi=\pi(\nu_2-\nu_1)t$ is
associated with the frequency difference of the two coherent
driving pump fields of the system.
\section{Appendix-B}
The expressions $P_{j=1-4}(\Delta)$ of the Eq. (3) are listed
bellow
\begin{eqnarray}
P_{1,2}(\Delta)=\frac{|\Omega_{1,2}|^2(2
\Gamma_{ad}[\Gamma_{ab}-i(\Delta-\delta_3)])}{M(\Delta)
(\Gamma_{ad}+\Gamma_{ac})(\Gamma^2_{ad}+\delta^2_{1,2})},
\end{eqnarray}
\begin{eqnarray}
P_3(\Delta)=\frac{|\Omega_1|^2 (K_1[\Gamma_{ab}-i(\Delta-\delta_1-\delta_3)]-\frac{\Omega^2_3}{4})}{M(\Delta)
(\Gamma_{ad}+i\delta_1)[{T_1}+\frac{\Omega^2_3}{4}]},
\end{eqnarray}
and
\begin{eqnarray}
P_4(\Delta)=\frac{|\Omega_2|^2 (K_1[\Gamma_{ab}-i(\Delta-\delta_2-\delta_3)]-\frac{\Omega^2_3}{4})}{M(\Delta)
(\Gamma_{ad}+i\delta_2)[{T_2}+\frac{\Omega^2_3}{4}]},
\end{eqnarray}
respectively, where
\begin{eqnarray}
M(\Delta)=(\Gamma_{ac}-i\Delta)(\Gamma_{ab}-i(\Delta-\delta_3))+\frac{\Omega^2_3}{4}.
\end{eqnarray}
However, in Eq. (6) $A_i$ for $i=1-4$ are given by
\begin{eqnarray}
A_{1,2}(kv,\Delta)=\frac{2\Gamma_{ad}(\Gamma_{ab}-i(2kv+\Delta-\delta_3))|\Omega_{1,2}|^2}{(\Gamma_{ad}+\Gamma_{ac})
(\Gamma^2_{ad}+(kv+\delta_{1,2})^2)[\beta+\frac{\Omega^2_3}{4}]},
\end{eqnarray}
\begin{eqnarray}
A_3(kv,\Delta)=\frac{[\alpha(\Gamma_{bd}-i(kv+\Delta-\delta_1-\delta_3))-\frac{\Omega^2_3}{4}]\Omega^2_1}{(\Gamma_{ad}+i(kv+\delta_1))[\beta+\frac{\Omega^2_3}{4}]B_1},
\end{eqnarray}
and
\begin{eqnarray}
A_4(kv,\Delta)=\frac{[\alpha(\Gamma_{bd}-i(kv+\Delta-\delta_2-\delta_3))-\frac{\Omega^2_3}{4}]\Omega^2_2}{(\Gamma_{ad}+i(kv+\delta_2))[\beta+\frac{\Omega^2_3}{4}]B_2},
\end{eqnarray}
respectively, where
$$B_{1,2}(kv,\Delta)=[(\Gamma_{cd}-i(\Delta-\delta_{1,2}))(\Gamma_{bd}-i(kv+\Delta-\delta_{1,2}-\delta_3))+\frac{\Omega^2_3}{4}],$$
$$T_{1,2}=(\Gamma_{cd}-i(\Delta-\delta_{1,2}))(\Gamma_{bd}-i(\Delta-\delta_{1,2}-\delta_3)),$$
$$\beta=(\Gamma_{ac}-i(kv+\Delta))(\Gamma_{ab}-i(2kv+\Delta-\delta_3)),$$
$$K_1=[\Gamma_{ab}-i(\Delta-\delta_3)],$$
and

$$\alpha=(\Gamma_{ab}-i(2kv+\Delta-\delta_3)).$$

\end{document}